\documentclass[10,fleqn]{article}
\usepackage{nccmath}
\usepackage[dvipdfmx]{graphicx}
\usepackage{wrapfig}
\usepackage{here}
\usepackage[top=3truecm, bottom=3truecm, left=2truecm, right=2truecm]{geometry}
\usepackage{amsmath,amssymb}
\usepackage{bm}
\usepackage{color}
\usepackage{ascmac}
\usepackage{array}
\usepackage{float}
\usepackage{authblk}
\usepackage{comment}
\usepackage{url}
\usepackage{lineno}
\usepackage[subrefformat=parens]{subcaption}
\title{\bf{Preparation and operation of\\ SiW-ECAL technological prototype\\ for DESY test beam 2019} }
\author{Kiichi Goto for CALICE SiW-ECAL group}
\affil{Department of Physics, Faculty of Science, Kyushu University, Fukuoka, Japan}
\date{}

\begin{document}
\maketitle
\begin{abstract}
We are developing SiW-ECAL technological prototypes with the aim of optimizing readout structure for $e^+\ e^-$ linear collider application.
Developments are being carried out under the cooperation of LAL, LLR and Kyushu University.
In this paper, we report the development status of FEV13, the latest version of the technological prototypes, following improvements from the previous versions.

Several beam tests have been conducted so far for this FEV13 performance test.
Most recently, beam test was conducted with electron beam (1-5 GeV) at DESY in Germany from June to July 2019.
We mainly report on the setup and various operation tests of FEV13 during this beam test.
\end{abstract}

\section{Introduction}

The International Linear Collider (ILC) is an $e^+\ e^-$ linear collider, to be constructed in Japan with the center-of-mass energy expected to be 250 GeV.
In the ILC, there are two candidates of multi-purpose detectors, the Silicon Detector (SiD) and the International Large Detector (ILD), and these detectors are designed with optimization for Particle Flow (PF) technique.
It is an algorithm with separation of the particles in jets and higher resolution of jet energy reconstruction can be obtained with optimized detectors.

In this paper, we show the development and preparation for the technological prototype of the silicon-tungsten electromagnetic calorimeter (SiW-ECAL) for DESY test beam in June-July 2019.
Related operation tests in Kyushu University are also reported.

\section{The SiW-ECAL for the ILD}

The electromagnetic calorimeter for the ILD is a sampling calorimeter with 20-30 layers.
In the SiW-ECAL, tungsten and silicon are used as the absorber and the detector, respectively.
The pixel size of the silicon sensor is 5 mm by 5 mm, which is decided by the requirement of PF algorithm.

\begin{figure}[H]
 \centering
 \includegraphics[width=5cm]{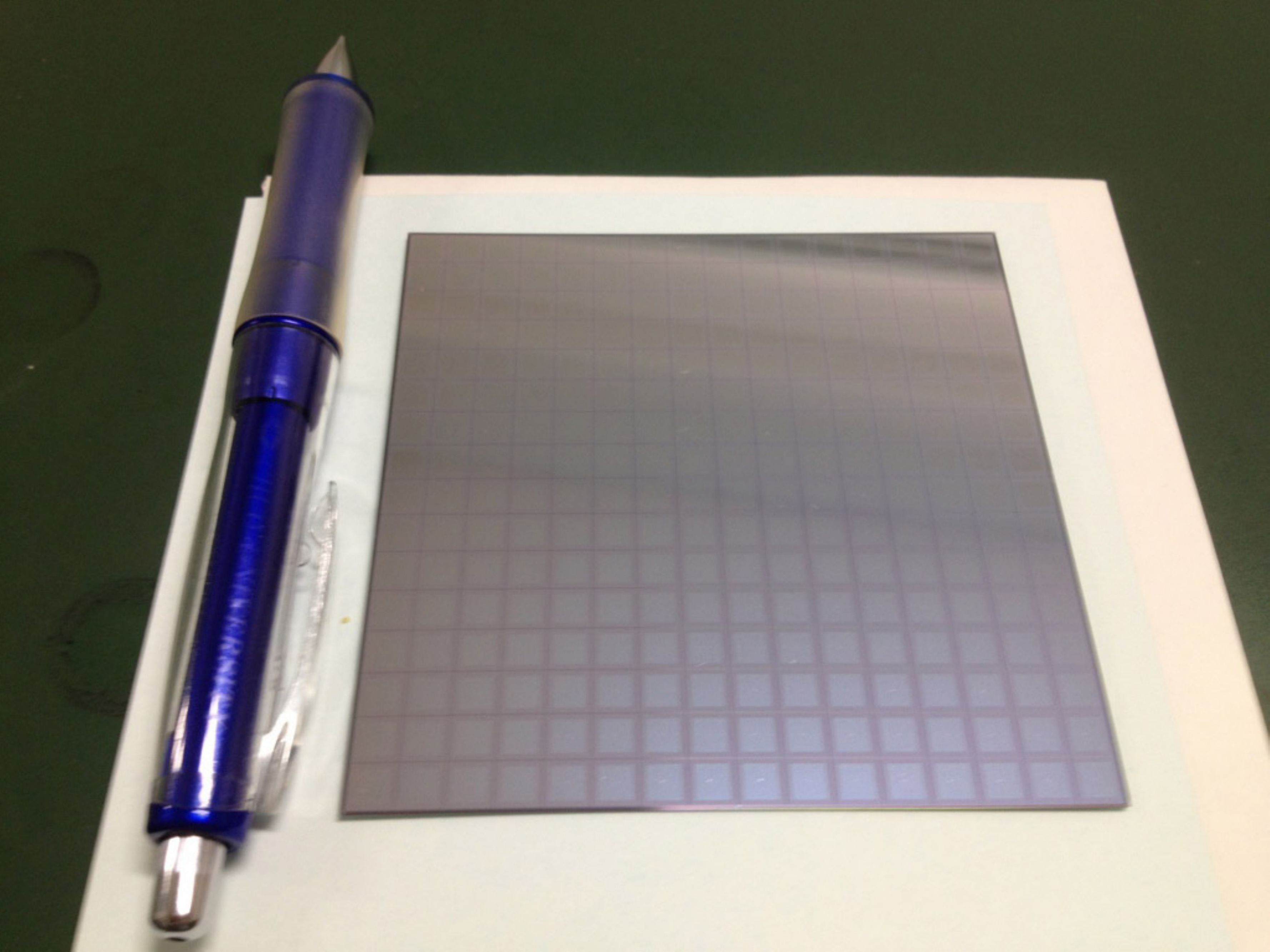}
 \caption{Silicon Pad 
 \protect\linebreak The size is 90 mm by 90 mm and the number of pixel is 256 (16 by 16).}
 \label{sensor}
\end{figure}

Actually, silicon pads shown in Figure \ref{sensor} are used.
The size of the silicon pad is 90 mm by 90 mm and the number of pixel is 256.
We use two patterns of the silicon thickness, 320 $\mathrm{\mu{}m}$ and 650 $\mathrm{\mu{}m}$.

\begin{figure}[H]
 \centering
 \includegraphics[width=14cm]{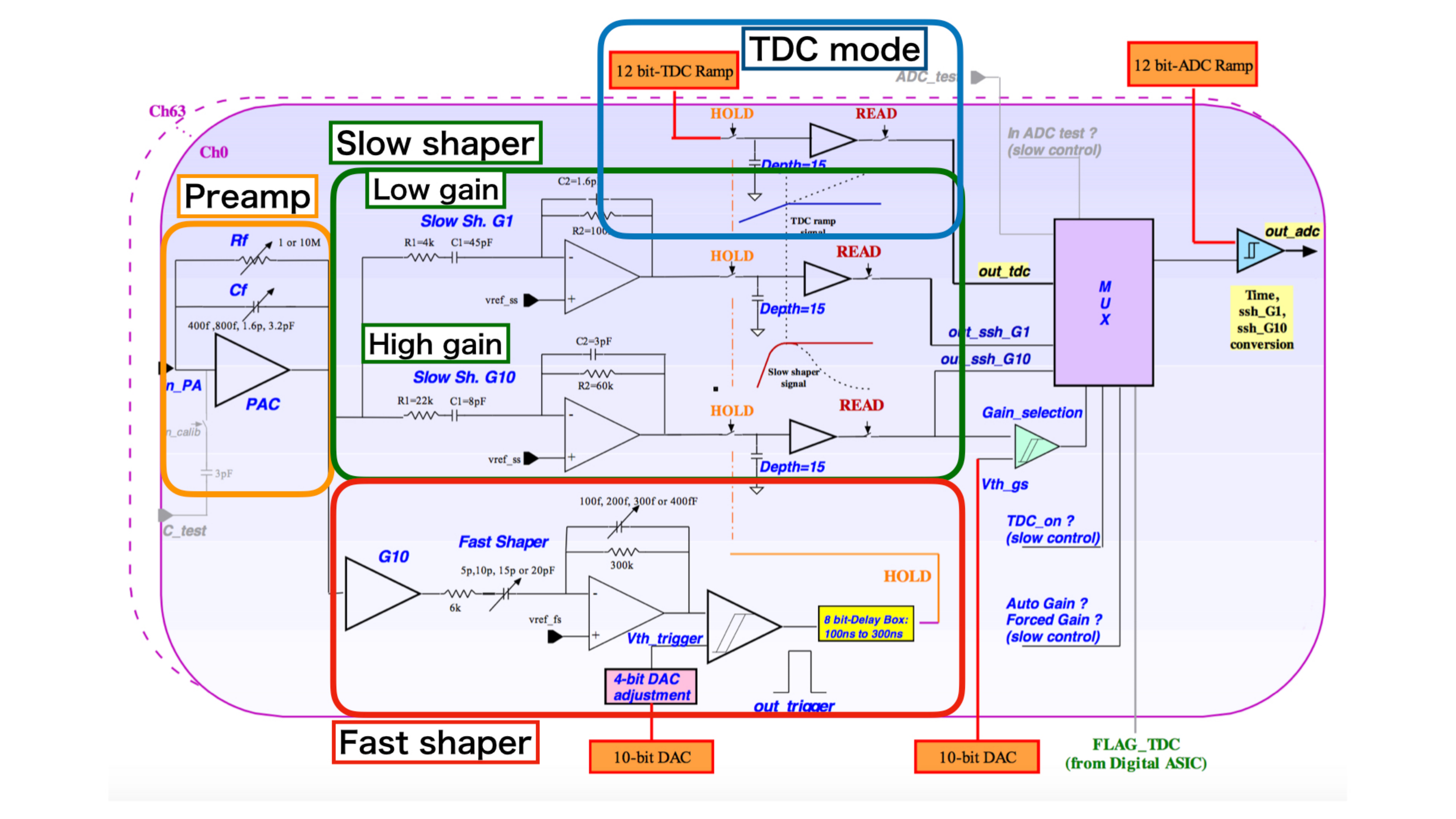}
 \caption{SKIROC2A 
 \protect\linebreak It contains preamp (orange), slow shaper (green) and fast shaper (red).}
 \label{asic}
\end{figure}

An Application-Specific Integrated Circuit (ASIC), SKIROC2A, dedicatedly developed for this prototype, are used for data taking.
SKIROC2A can acquire 64 channels and hold charges in 15 memory cells per channel.
The signal entering the circuit, first, passes through a preamplifier (labeled as Preamp in Figure \ref{asic}), then processed by two shapers.
The self trigger is provided by the fast shaper, and the signal is reshaped for charge measurements by the slow shaper.
The Slow shaper contains two gains, high gain and low gain with the 10 times, difference between the two gains.
In addition, SKIROC2A has two readout modes, ADC and TDC.
In the ADC mode, we can take two gains, high gain and low gain, as the readout data.
In the TDC mode, TDC data can be taken and we can also choose one gain of three ADC gains, high, low or auto gain.
Auto gain is the automatically switching mode between high and low gain when the signal exceed a certain threshold.

\begin{figure}[H]
 \centering
 \includegraphics[width=8cm]{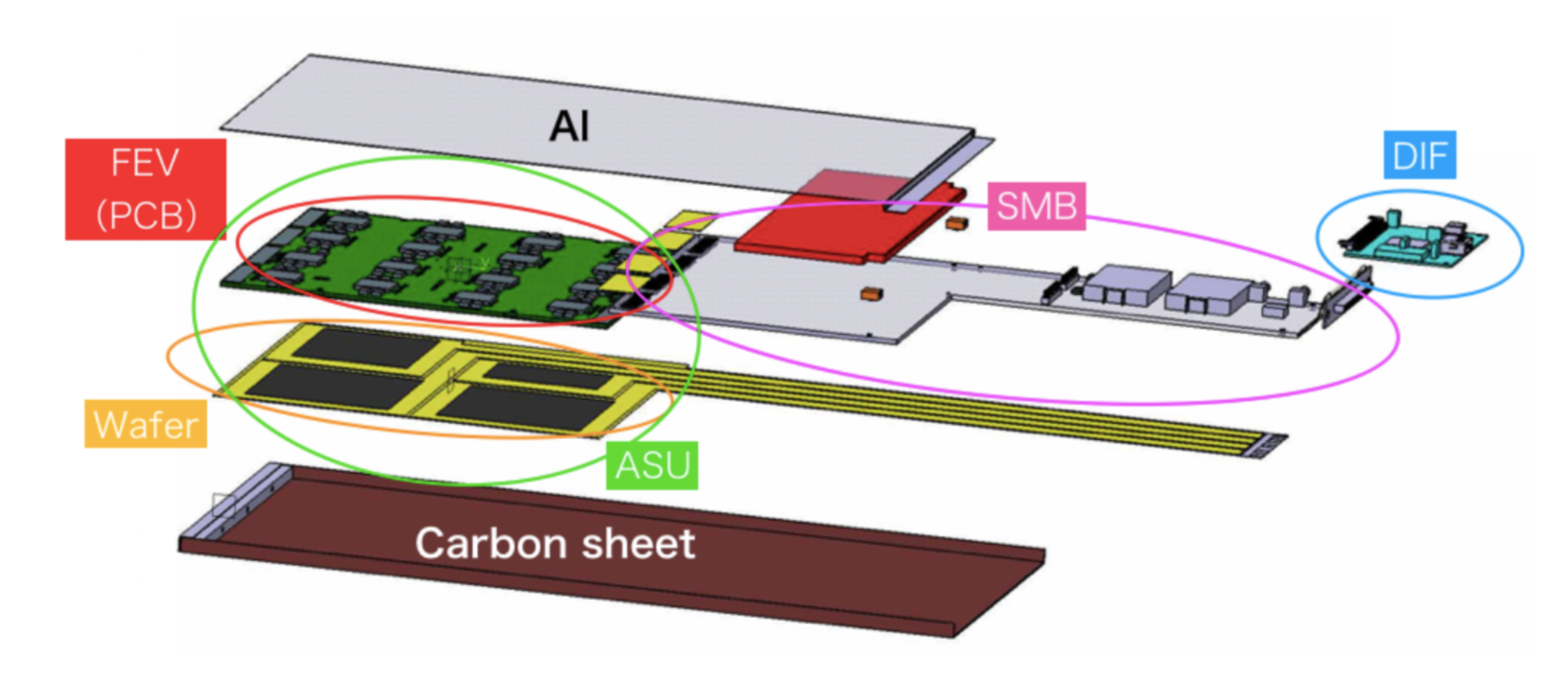}
 \caption{Slab component (FEV11)}
 \label{slab_component}
\end{figure}

One layer of the detector complex is called a ``slab (short slab)".
A slab contains four silicon pads and these are read out by 16 ASIC, shown in Figure \ref{slab_component}.
The DIF is used for data transmission and receiving and control of the clock.
The SMB is used as data storage and power supplier to the ASICs.

Now we are studying the technological prototype, ``FEV13".
Improvements from the previous model (FEV11) are shown below.
\begin{itemize}
  \item Replacement of ASIC, from SKIROC2 to SKIROC2A, for improvements on TDC and individual threshold control
  \item Change the connection of between FEV and SMB, from a soldered kapton sheet to cables
  \item Smaller footprint of SMB
\end{itemize}

\begin{figure}[H]
 \centering
 \includegraphics[width=12cm]{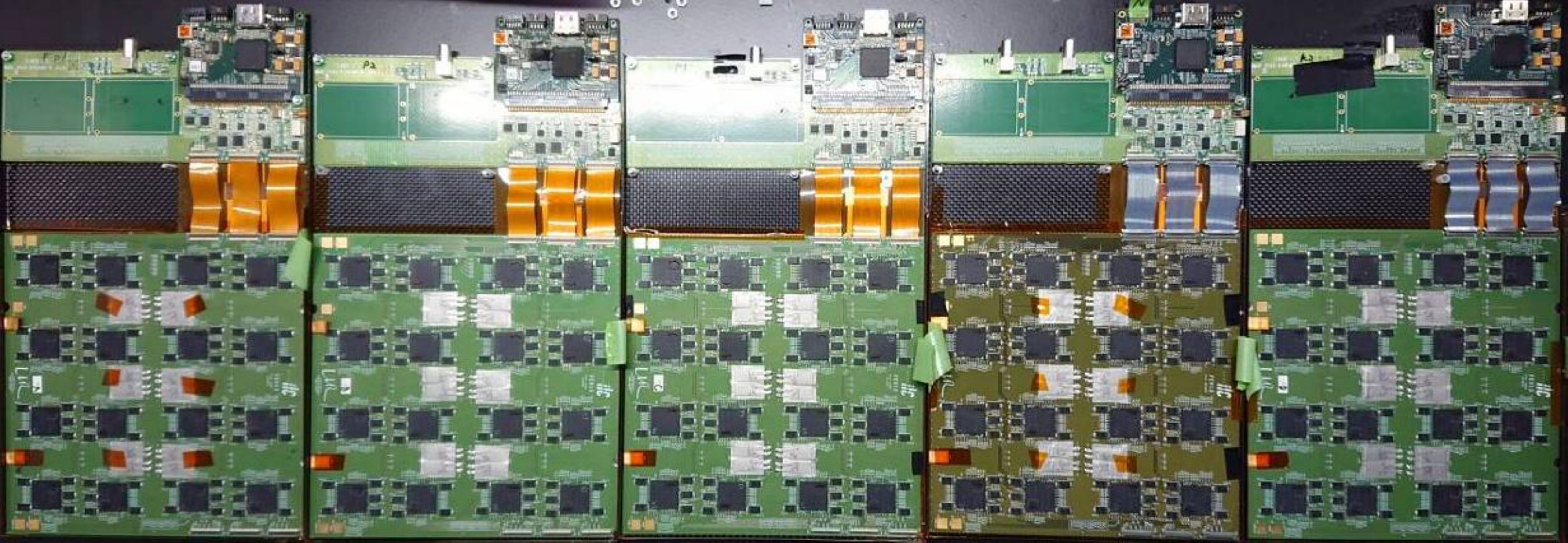}
 \caption{All slabs (FEV13) 
 \protect\linebreak assembled in Kyushu University
 \protect\linebreak named from left as P1, P2, P3, K1 and K2}
 \label{slabs}
\end{figure}

All slabs assembled in Kyushu University are shown in Figure \ref{slabs}.
The silicon thickness of slab P3 is 320 $\mathrm{\mu{}m}$, and the thickness of other slabs is 650 $\mathrm{\mu{}m}$.
The difference between P and K type is cables connecting SMBs to FEVs.
We use flexible cables in the P type and we use micro-coaxial cables in the K type.

\section{Preparation for DESY test beam 2019}

We performed two test beams for FEV13 before the test beam in 2019.
The first test beam was performed at DESY in 2018, and just one slab was tested with electron beam.
In this test beam, we used a rigid printed circuit board (PCB) for the high voltage (HV) connection.
The second test beam was performed at CERN/SPS in 2018.
We evaluated five slabs with electron, muon and pion beams and the HV connection was replaced to flex circuits.
In this test beam, four slabs of five had a problem in HV connection due to reused carbon plates of an older model (FEV8).
Therefore, we couldn't take data well.

The test beam at DESY in 2019 was the third test beam for FEV13.
After the previous test beam, we replaced the carbon plates to new ones.
New carbon plate has following improvement points and problems.
In the new carbon plates we added an edge cut to accommodate HV connectors and a screw hole to push plates to the flex circuit for HV.
These modifications helped to prevent the HV disconnection issue, while the new plates also introduce a new issue that the HV limit sets lower due to imperfect insulation of the plates.

In this test beam we introduced temperature monitoring by a chip on the SMB.
The temperature sensors were already put on the SMBs in the previous test beam, but temperature measurements were not done since the feature had not been implemented in the firmware.

\section{Operation test in the Kyushu University}

We performed some operation tests for this test beam.
First, pedestal means are measured with thermal noise.
Second, we confirmed ADC linearity with two radiation sources.
Finally, we calculated relation between TDC and real time with charge injection.

\subsection{Measurement of pedestal with thermal noise}

\begin{figure}[H]
 \centering
 \includegraphics[width=14cm]{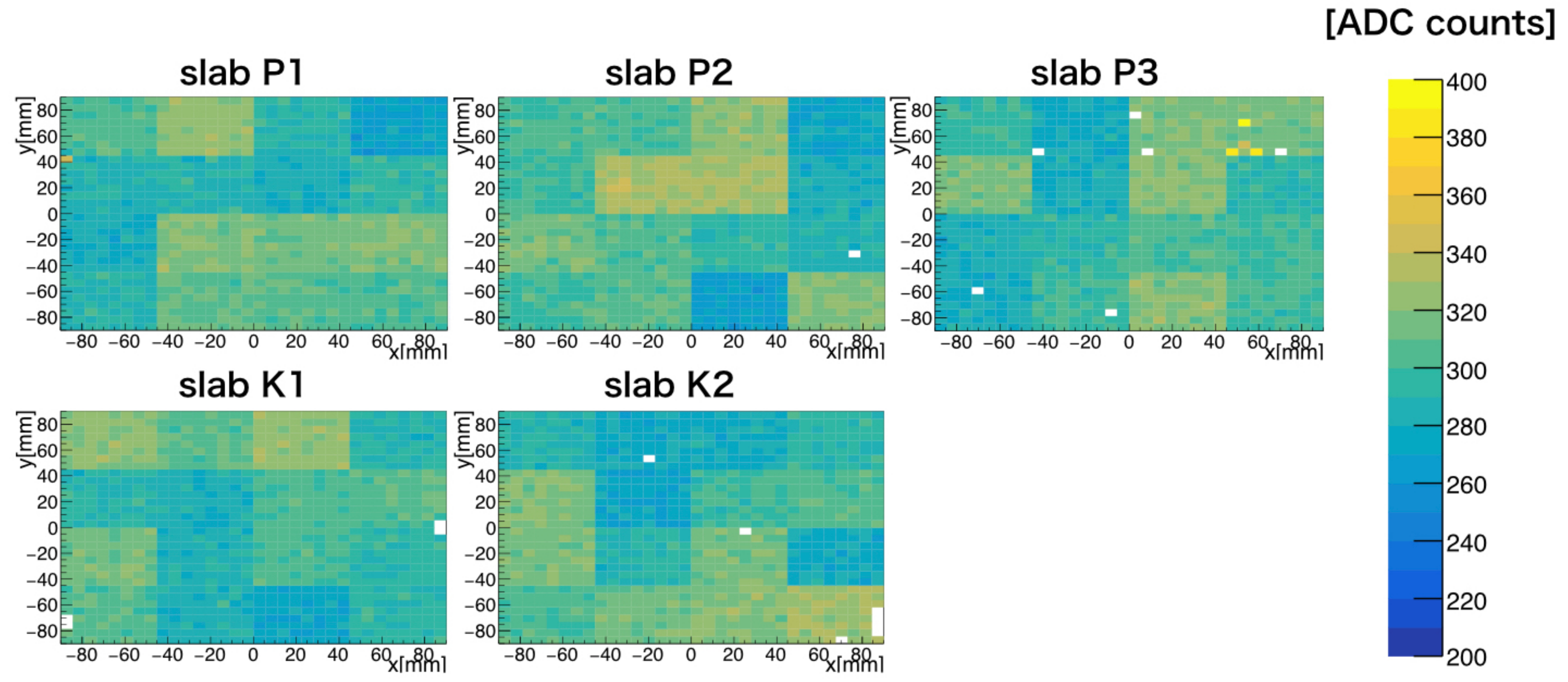}
 \caption{Pedestal Map\\The color scale is ADC counts.}
 \label{pedestal_map}
\end{figure}

Pedestal means for all channels in the first memory cell using high gain are shown in Figure \ref{pedestal_map}.
The data of all 64 channels are recorded when at least one channel is triggered.
In this pedestal analysis, we use the non-triggered data with bad data removed.
The pedestal means are almost the same within each ASIC chip and we can confirm that they are stable within it.
We can see that some channels are blacked.
This is expected that these channels have very low statistics, in other words, they have much higher triggered rate (noisy channels), since we request the data of non-triggered channels in the pedestal analysis.
Therefore we cannot measure the pedestal in channels with higher triggered rate.

\subsection{ADC measurement with radiation sources}

We took ADC data with $^{133}\mathrm{Ba}$ and $^{57}\mathrm{Co}$ as gamma radiation sources.
$^{133}\mathrm{Ba}$ emits gamma ray of 356 keV (62.1\%) and 81.0 keV (34.1\%), but we cannot see 356 keV photoelectric peak because of the samll cross section.
We can see the Compton edge at 207 keV instead in this measurement.
$^{57}\mathrm{Co}$ emits gamma ray of 122 keV (85.6\%) and 136 keV (10.7\%).
Only 122 keV peak can be seen in this measurement.

\begin{figure}[H]
 \centering
 \includegraphics[width=9cm]{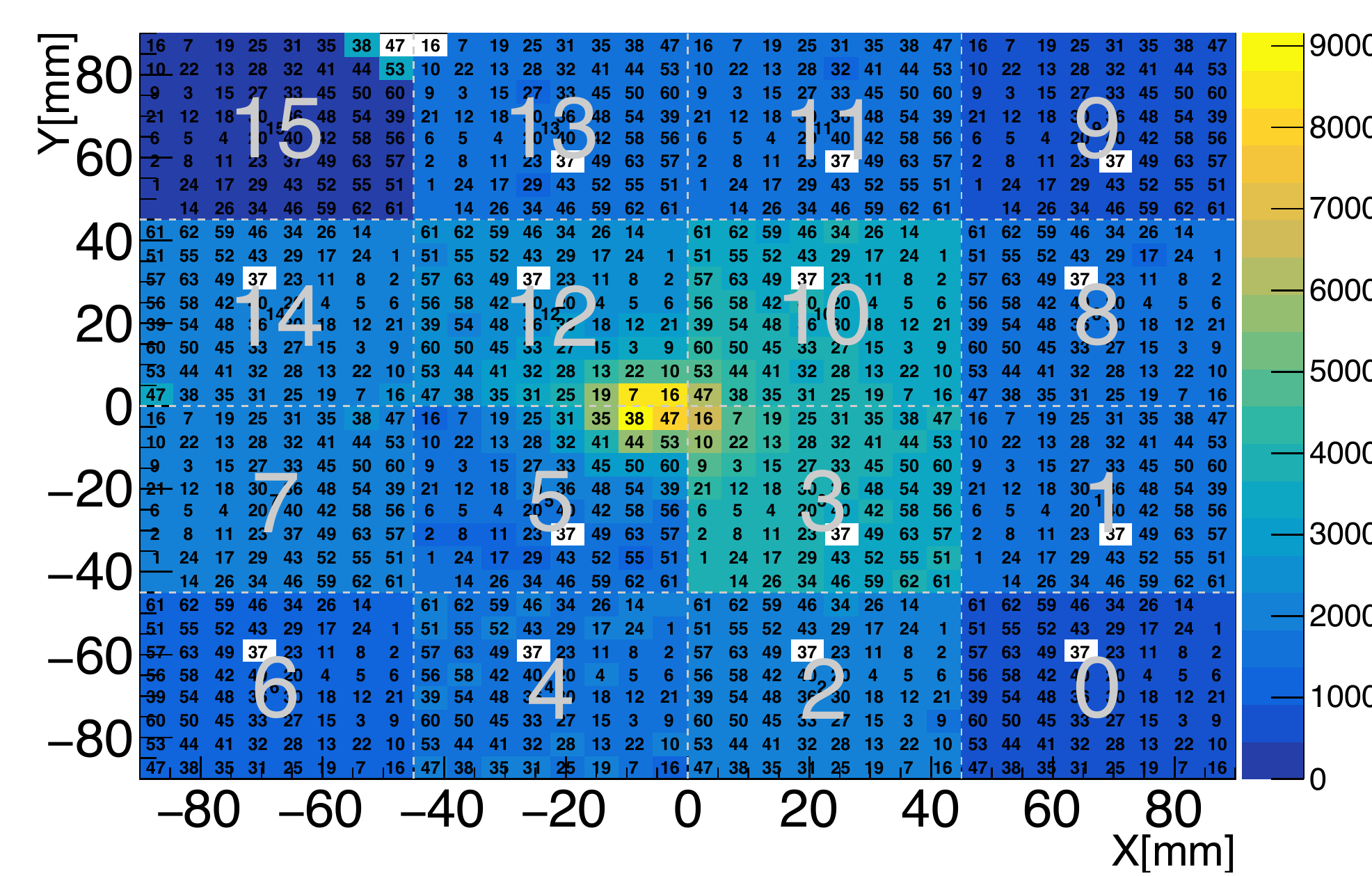}
 \caption{Hit map (Ba 133) 
 \protect\linebreak The color scale is the number of hit.}
 \label{ba_hitmap}
\end{figure}

The hit map of $^{133}\mathrm{Ba}$ is shown in Figure \ref{ba_hitmap}.
The hit map shows number of the triggered events at each cell, with $^{133}\mathrm{Ba}$ source on the center of the slab.

\subsubsection{$^{133}\mathrm{Ba}$}

An ADC distribution of $^{133}\mathrm{Ba}$ after the pedestal subtraction is shown in Figure 7(a), with logarithmic scale as the vertical axis.
The peak around zero ADC counts is the pedestal peak.
We can see one photoelectric peak and one Compton edge.
The photoelectric peak of 81.0 keV was fitted using Gaussian.
We can obtain 64.8 ADC counts as the mean ADC value of the photoelectric peak.
The Compton edge of 207 keV was fitted using an error function.
We can obtain 173 ADC counts as the ADC value of the Compton edge.

\subsubsection{$^{57}\mathrm{Co}$}

An ADC distribution of $^{57}\mathrm{Co}$ after the pedestal subtraction is shown in Figure 7(b), with log scale as the vertical axis and the peak around zero ADC counts is the pedestal peak.
We can see two peaks.
I expect that one peak around 30 ADC counts is probably noise from pedestal.
The other is the photoelectric peak of 122 keV and I fitted this peak with Gaussian.
We can obtain 97.9 ADC counts as the mean ADC value of the photoelectric peak.

\begin{figure}[htbp]
  \begin{tabular}{cc}
   \begin{minipage}{0.5\hsize}
     \centering
     \includegraphics[width=8cm]{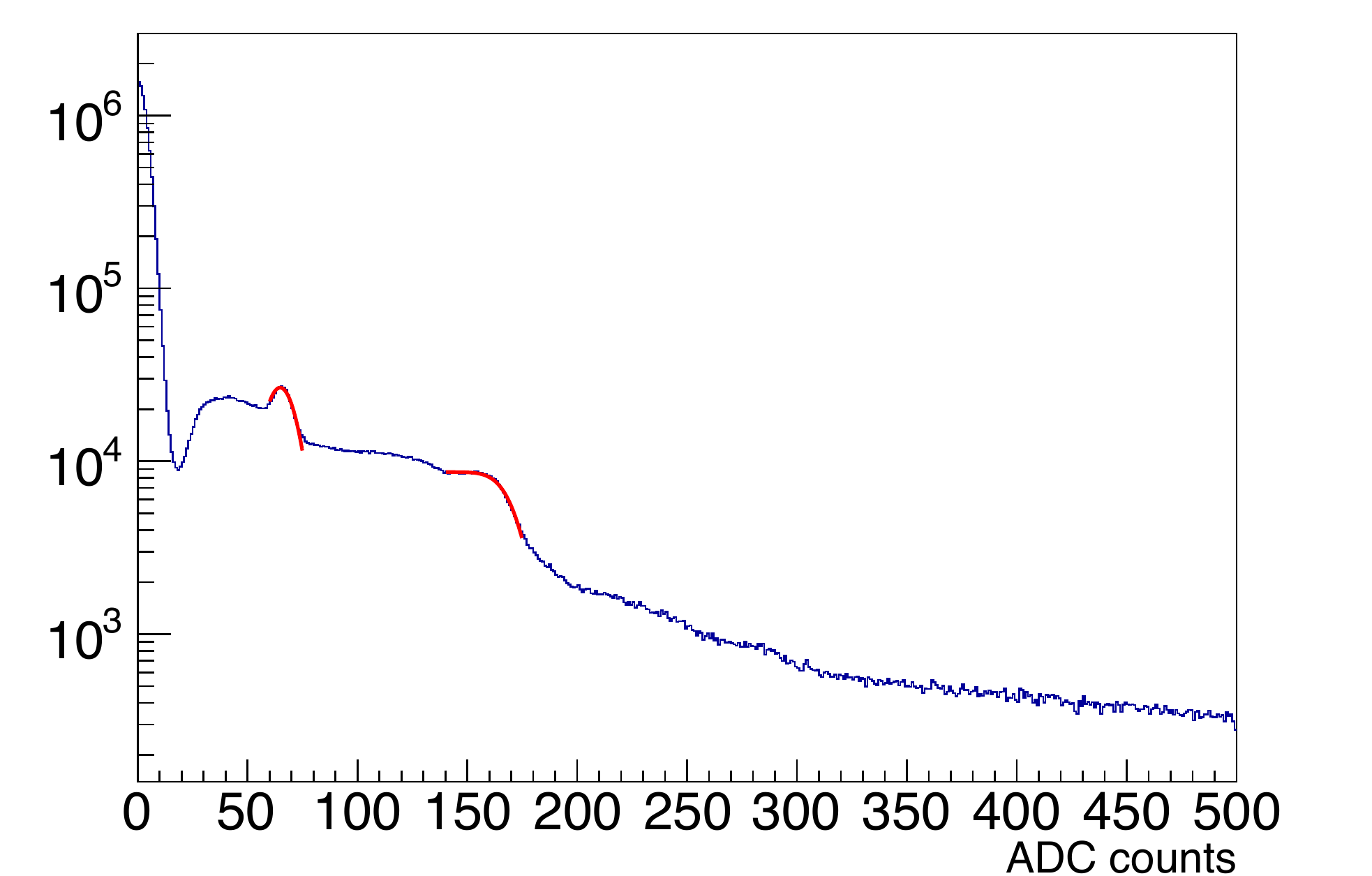}
     \subcaption{$^{133}\mathrm{Ba}$ 
     \protect\linebreak blue : $^{133}\mathrm{Ba}$ histogram 
     \protect\linebreak red : fitting function}
     \label{ba_hist}
   \end{minipage}
 \end{tabular}
 \begin{tabular}{cc}
   \begin{minipage}{0.5\hsize}
     \centering
     \includegraphics[width=8cm]{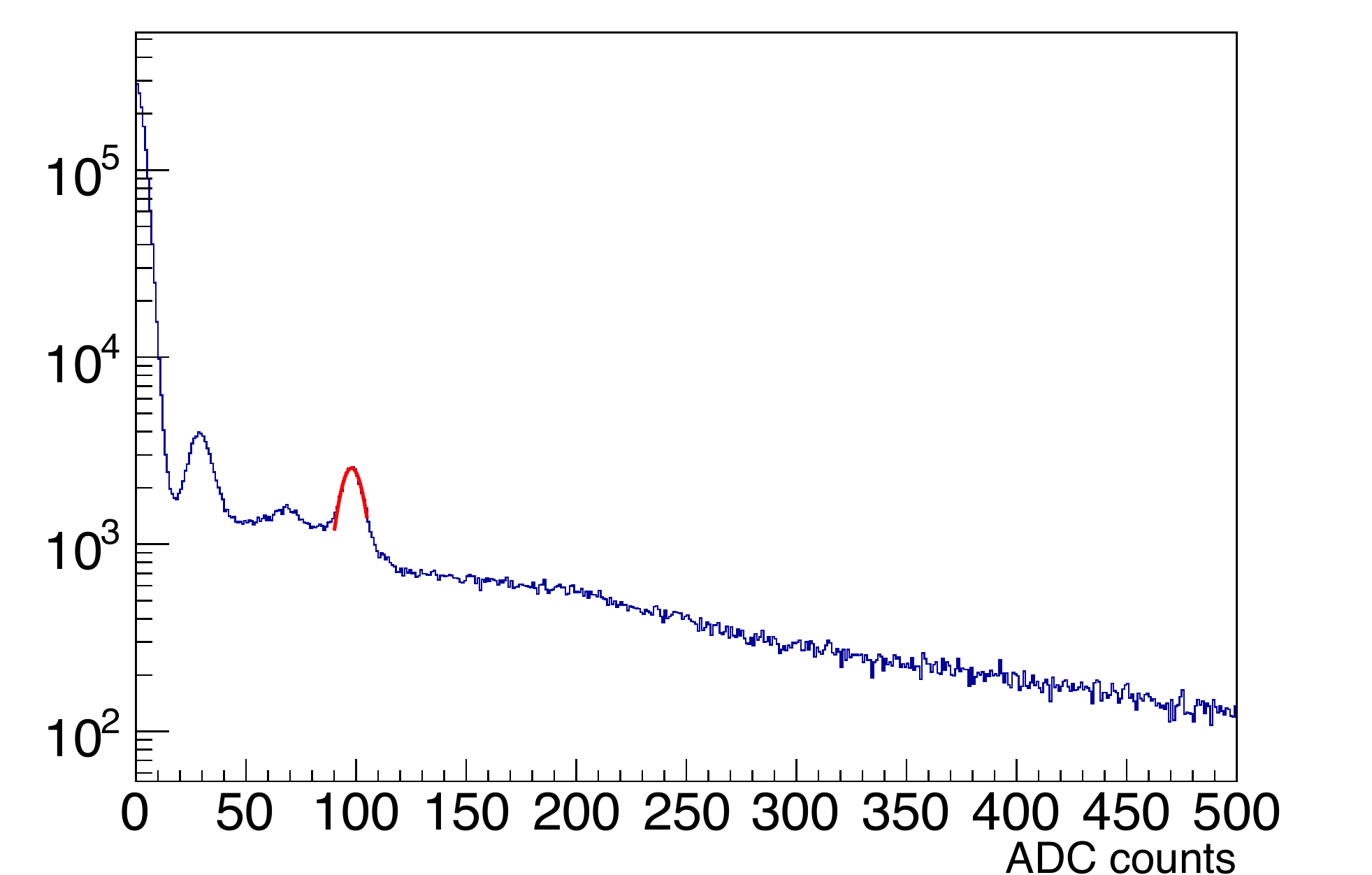}
     \subcaption{$^{57}\mathrm{Co}$
     \protect\linebreak blue : $^{57}\mathrm{Co}$ histogram 
     \protect\linebreak red : fitting function}
     \label{co_hist}
   \end{minipage}
 \end{tabular}
 \caption{ADC histogram after the pedestal subtracted}
\end{figure}

\newpage
\subsubsection{ADC linearity}

\begin{figure}[H]
 \centering
 \includegraphics[width=8cm]{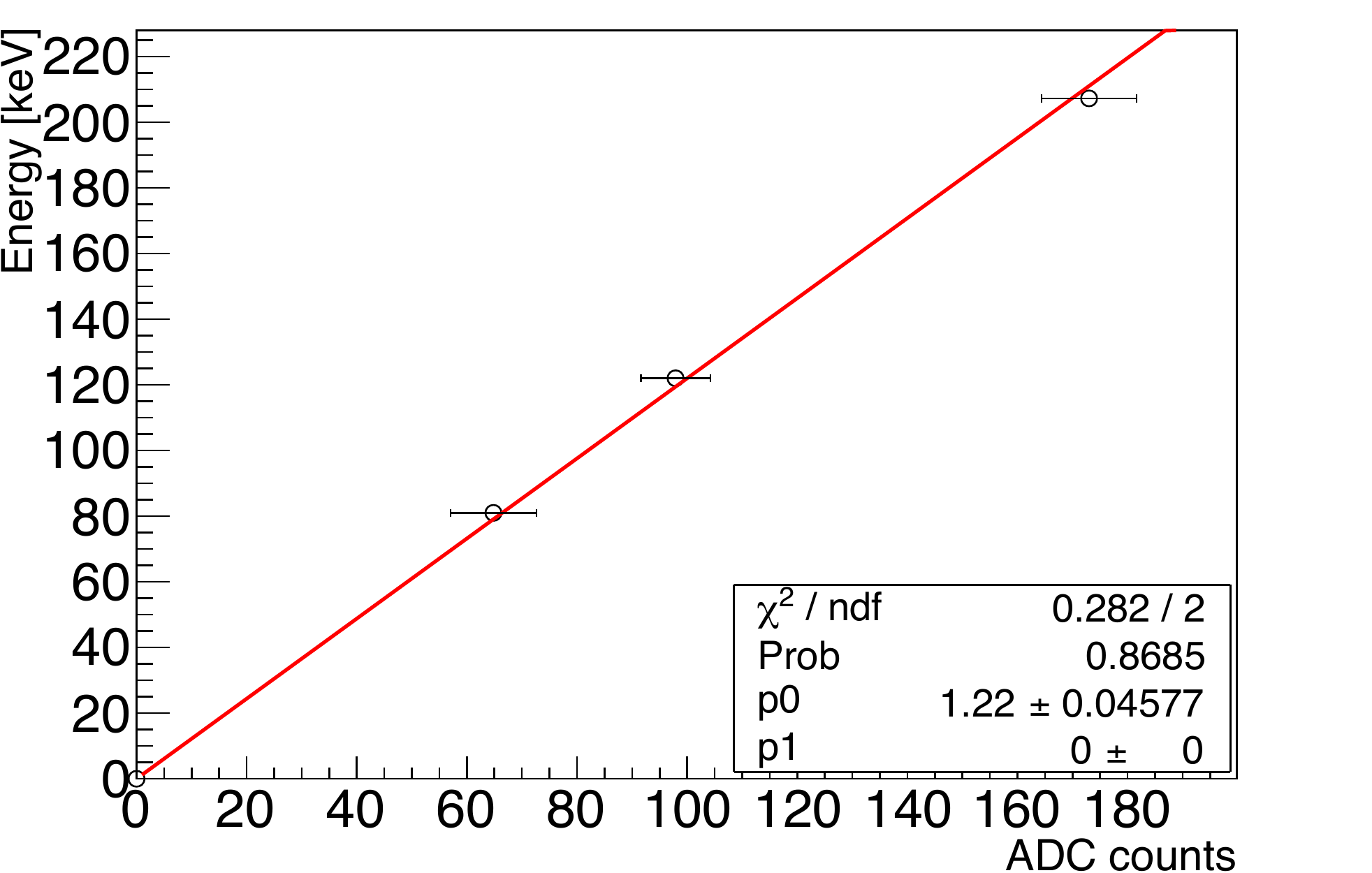}
 \caption{ADC linearity}
 \label{adc_linearity}
\end{figure}

Linearity of ADC counts and energy was checked with $^{133}\mathrm{Ba}$ and $^{57}\mathrm{Co}$ data (Figure \ref{adc_linearity}).
The fitting function is fixed passing through an origin.
The horizontal axis and the vertical axis are ADC value and energy, respectively.
Good linearity up to 200 keV is seen.

\subsection{TDC measurement with charge injection}

We performed a TDC operation test with charge injection.
A clock of 5 MHz is provided to SKIROC2A for bunch counting and TDC measurement.
A ramp waveform is produced with the clock for TDC measurement.
In Figure 9(a), the ramp waveform is shown.
The yellow line and red line are the TDC ramp waveform and the clock waveform respectively.
A scale on the horizontal axis is 100 ns and a scale on the vertical axis is 500 mV.
As shown, the frequency of the ramp waveform is the half of the clock in order to ensure the one-by-one relation between the TDC value and the real time.
Because of this feature, we have to treat odd and even bunch crossing separately.

\begin{figure}[htbp]
  \begin{tabular}{cc}
   \begin{minipage}{0.3\hsize}
     \centering
     \includegraphics[width=5cm]{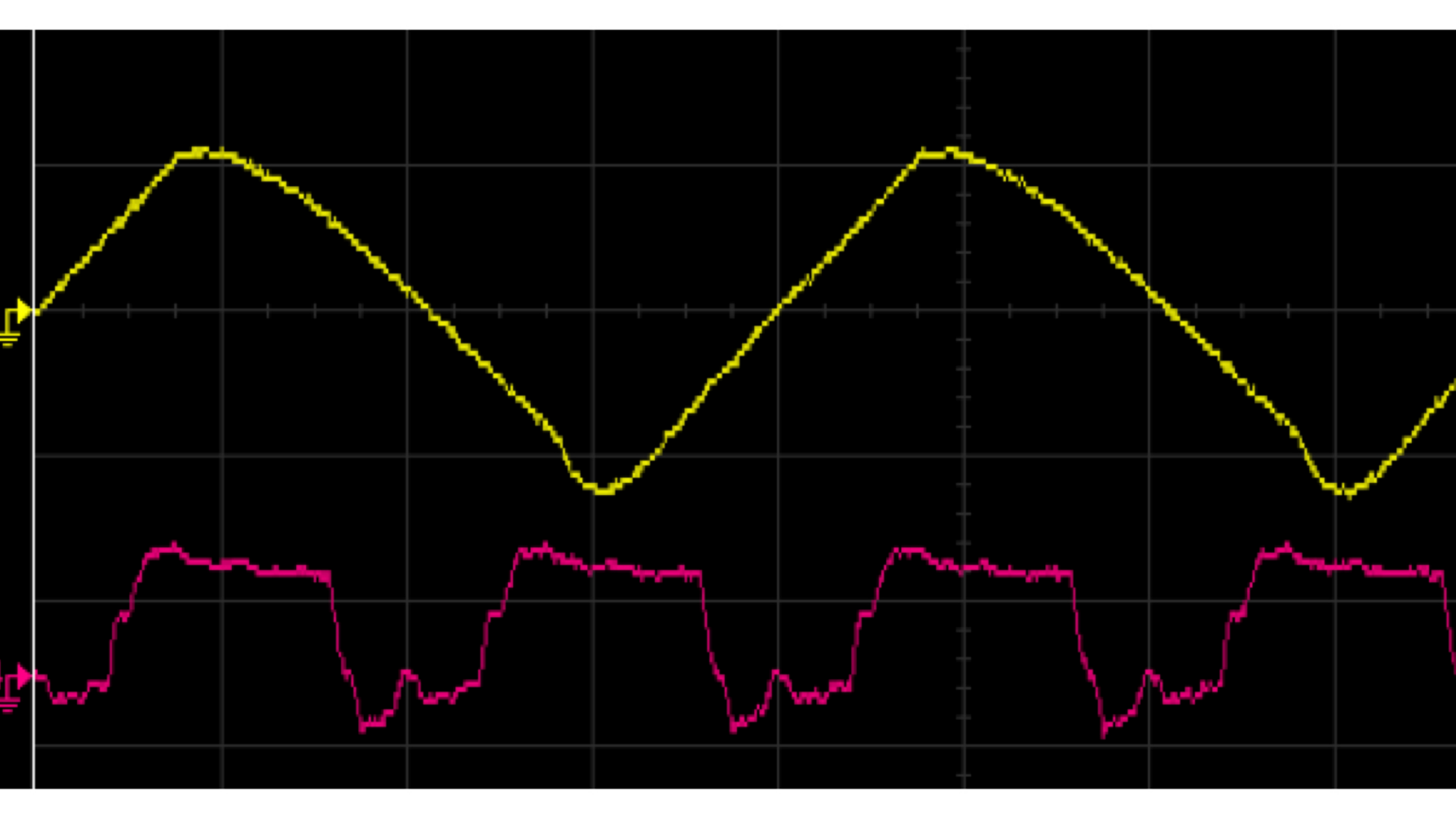}
     \subcaption{Inner ramp waveform  \cite{sekiya}
     \protect\\ yellow : ramp waveform
     \protect\\ red : clock signal}
     \label{oscilloscope_ramp}
   \end{minipage}
  \end{tabular}
  \begin{tabular}{cc}
   \begin{minipage}{0.3\hsize}
     \centering
     \includegraphics[width=5cm]{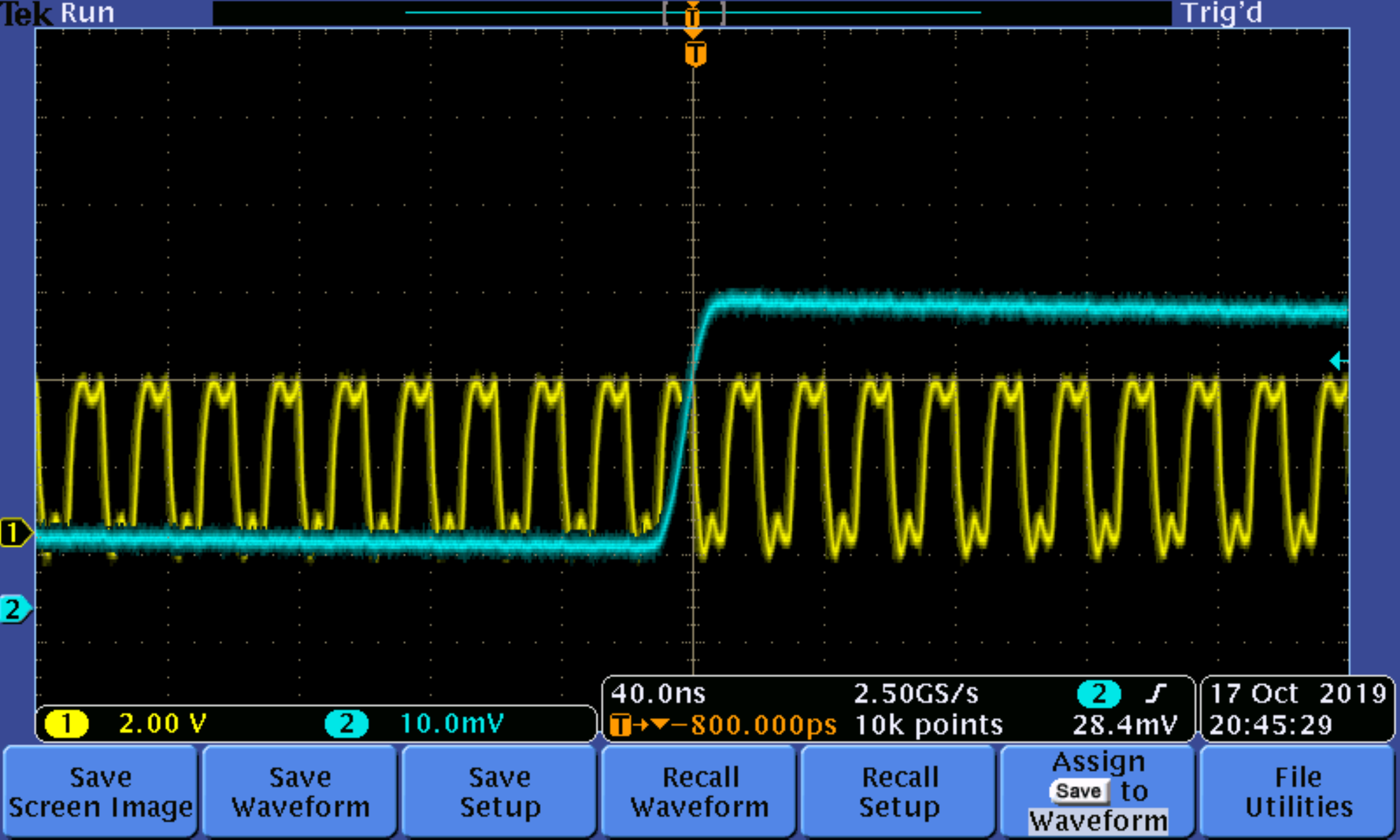}
     \subcaption{synchronized
     \protect\\ yellow : clock signal
     \protect\\ blue : injection signal}
     \label{oscilloscope_s}
   \end{minipage}
 \end{tabular}
 \begin{tabular}{cc}
   \begin{minipage}{0.3\hsize}
     \centering
     \includegraphics[width=5cm]{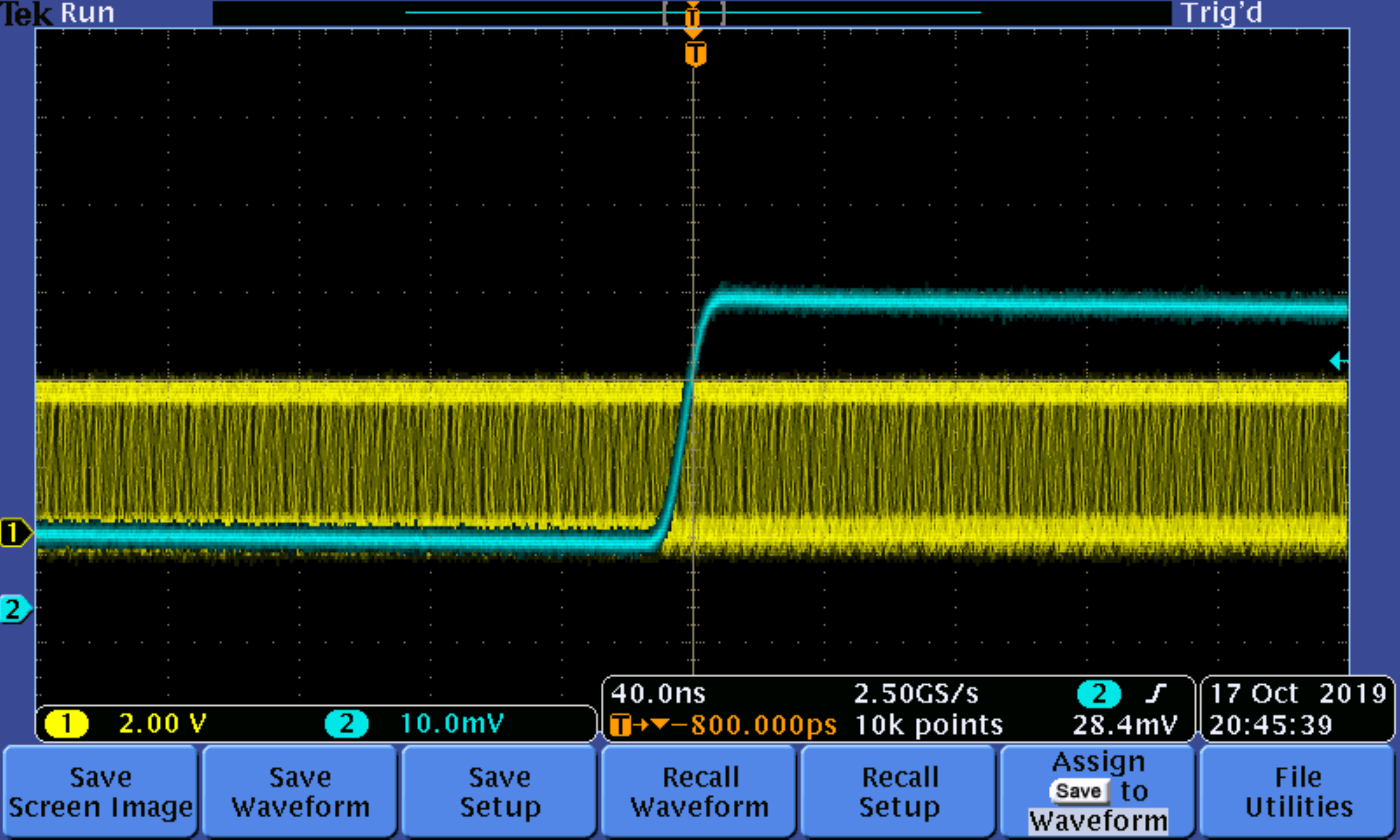}
     \subcaption{not synchronized
     \protect\\ yellow : clock signal
     \protect\\ blue : injection signal}
     \label{oscilloscope_ns}
   \end{minipage}
 \end{tabular}
 \caption{Monitoring on the oscilloscope}
\end{figure}

\begin{figure}[H]
 \centering
 \includegraphics[width=10cm]{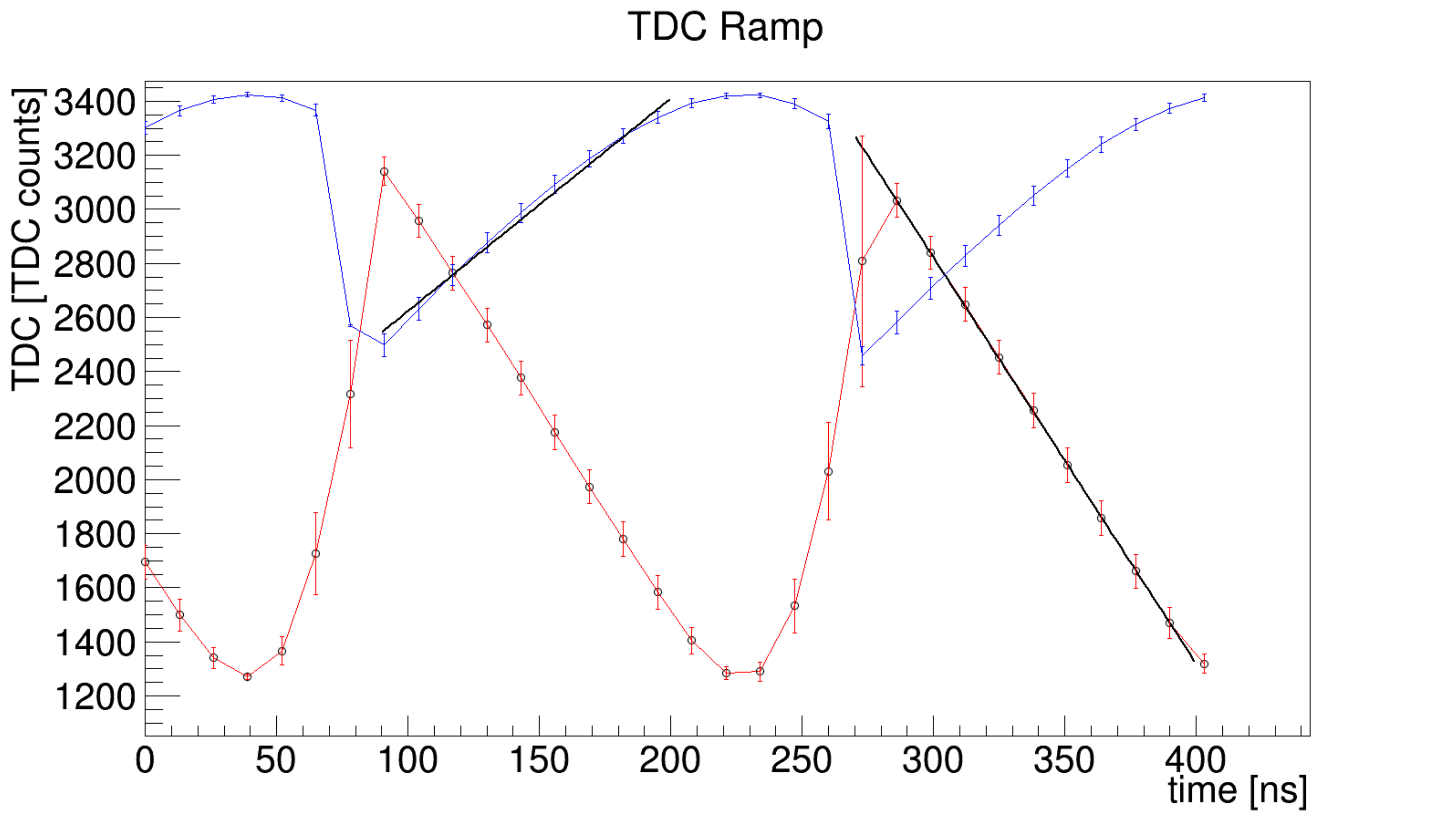}
 \caption{Inner ramp waveform
 \protect\linebreak red : bunch crossing even
 \protect\linebreak blue : bunch crossing odd}
 \label{tdc_ramp}
\end{figure}

To calculate the factor between TDC value and the real time, the TDC ramp waveform is measured by changing the phase of the injection signal synchronized with the clock.
Figure 9(b) shows a state that the signals are synchronized.
Figure 9(c) shows a state that the signals are not synchronized.
In this measurement, I used following parameters.
\begin{itemize}
  \item Bunch crossing clock : 5 MHz
  \item Injection frequency : 200 kHz
  \item Injection charge : 8.4 fC (equivalent to 1 MIP with 650 $\mathrm{\mu{}m}$ silicon)
\end{itemize}
The result of this measurement is shown in Figure \ref{tdc_ramp}.
We get two waveforms, red line and blue line, corresponding to the even and odd bunch crossing.
We fitted these lines and we could obtained the TDC to real time factors (Table 1(a)).
This values are comparable to the previous study without sensors (Table 1(b)).

\begin{table}[h]
 \caption{TDC to real time factor}
 \begin{tabular}{cc}
  \begin{minipage}{0.5\hsize}
   \subcaption{This study}
   \centering
    \begin{tabular}{|c|c|} \hline
    Bunch crossing & TDC factor \\ \hline
    Even (Down) & 0.127 ns/TDC count \\ \hline
    Odd (Up) & 0.066 ns/TDC count \\ \hline
    \end{tabular}
    \label{tdc_factor}
  \end{minipage}
  \begin{minipage}{0.5\hsize}
   \subcaption{Previous study  \cite{sekiya}}
   \centering
    \begin{tabular}{|c|c|} \hline
    Bunch crossing & TDC factor \\ \hline
    Down & 0.080 ns/TDC count \\ \hline
    Up & 0.094 ns/TDC count \\ \hline
    \end{tabular}
    \label{pre_factor}
  \end{minipage}
 \end{tabular}
\end{table}

\newpage
\section{DESY test beam 2019}

We performed a test beam at DESY during $24^{th}$ June to $7^{th}$ July 2019.
We used the electron beam and the beam energy is 1 to 5 GeV.
We use the 3 GeV beam energy for most of the measurements.
Five slabs and four Chip-in-Boards(CIB) were brought from Kyushu University and French group, respectively.
Chip-in-Boards are trial boards under development.
We performed the test beam using 9 slabs in total.

\begin{figure}[H]
 \centering
 \includegraphics[width=7cm]{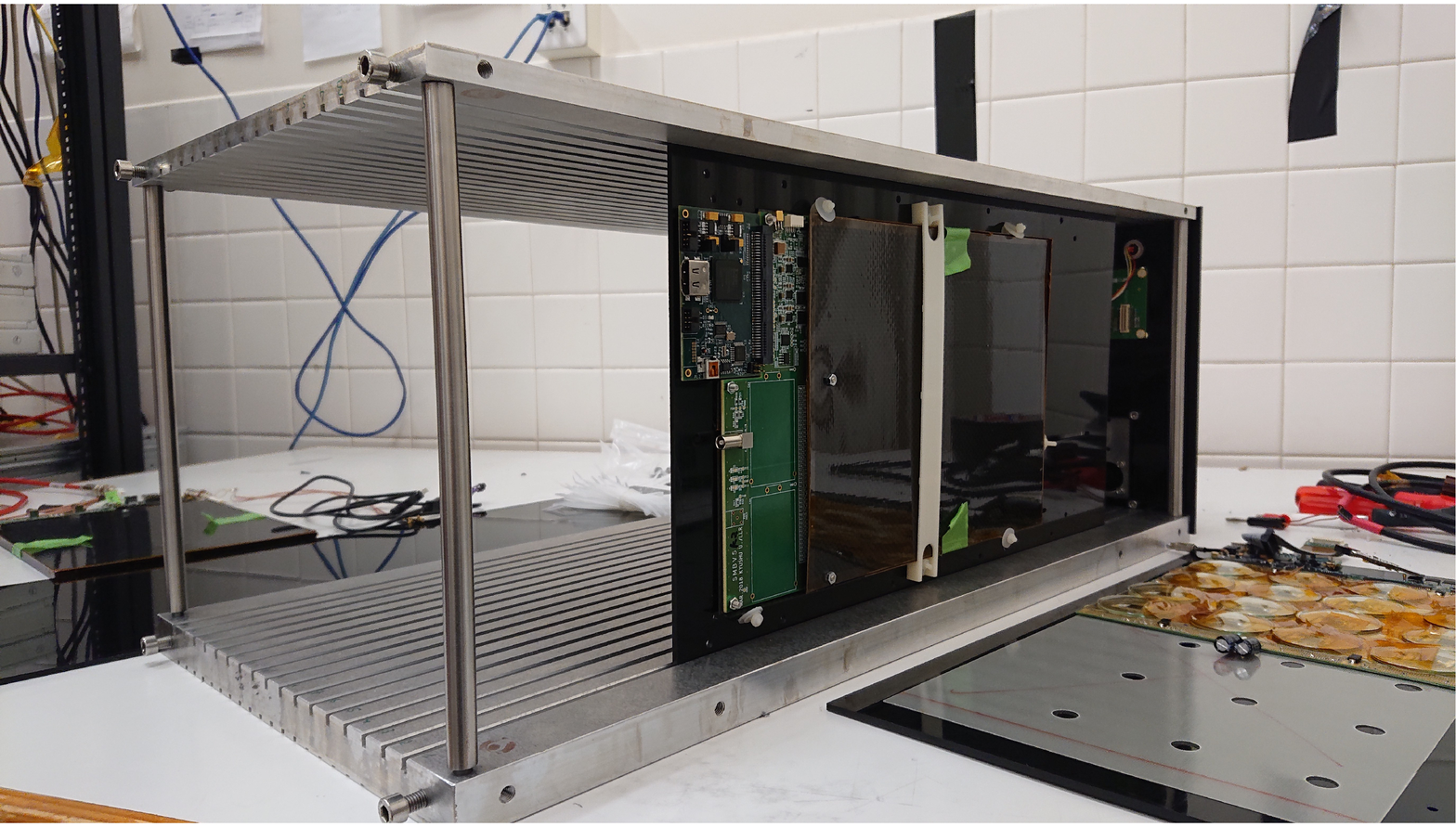}
 \caption{The box used in the test beam 2019}
 \label{desy_setup}
\end{figure}

In this test beam, we used two setups, MIP setup and shower setup.
Actually, the MIP setup is the state without tungsten absorber and the shower setup is the state with tungsten.
We used the box (Figure \ref{desy_setup}) for all the measurements and all slabs are installed to this box.
The order of the slabs is from the beam side, P1, P2, P3, K1, K2.
In the shower setup, we used two types of thicknesses of tungsten plates, 2.1 mm and 4.2 mm.
\\
\\

\begin{figure}[H]
 \centering
 \includegraphics[width=12cm]{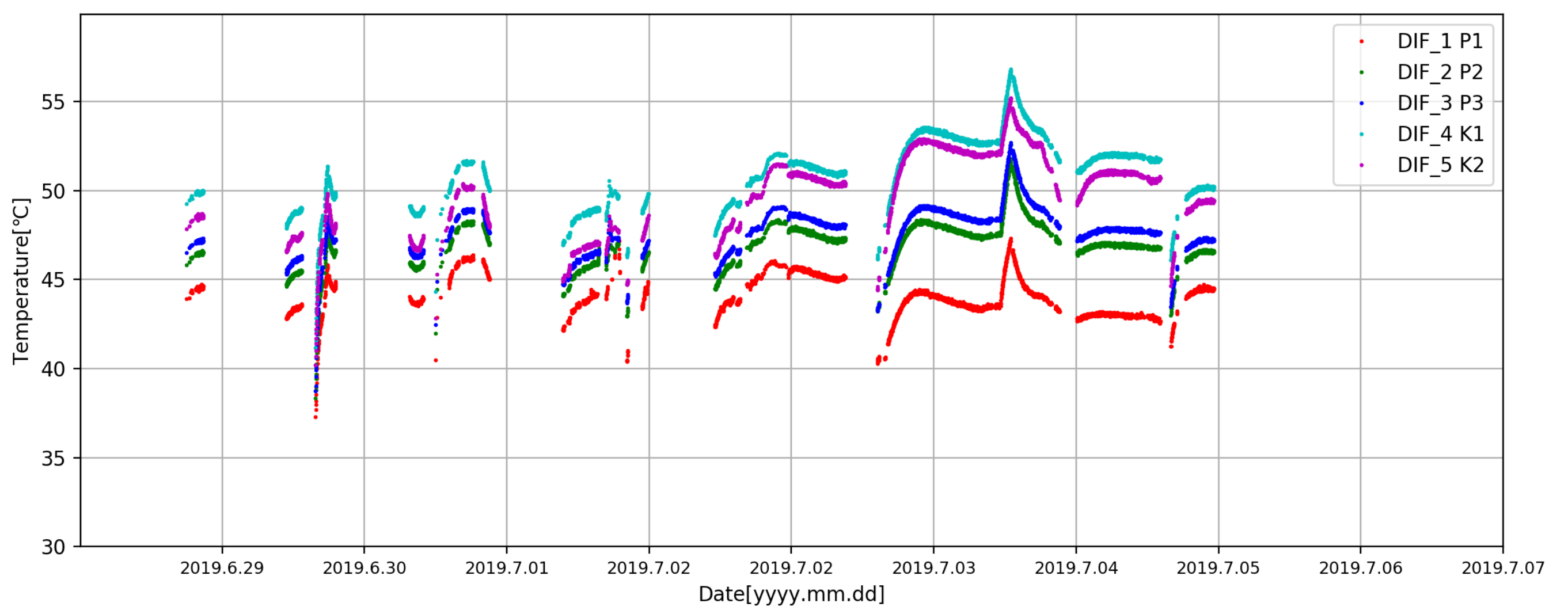}
 \caption{Time dependance of temperature for all slabs}
 \label{slab_temp}
\end{figure}

\begin{figure}[H]
 \centering
 \includegraphics[width=14cm]{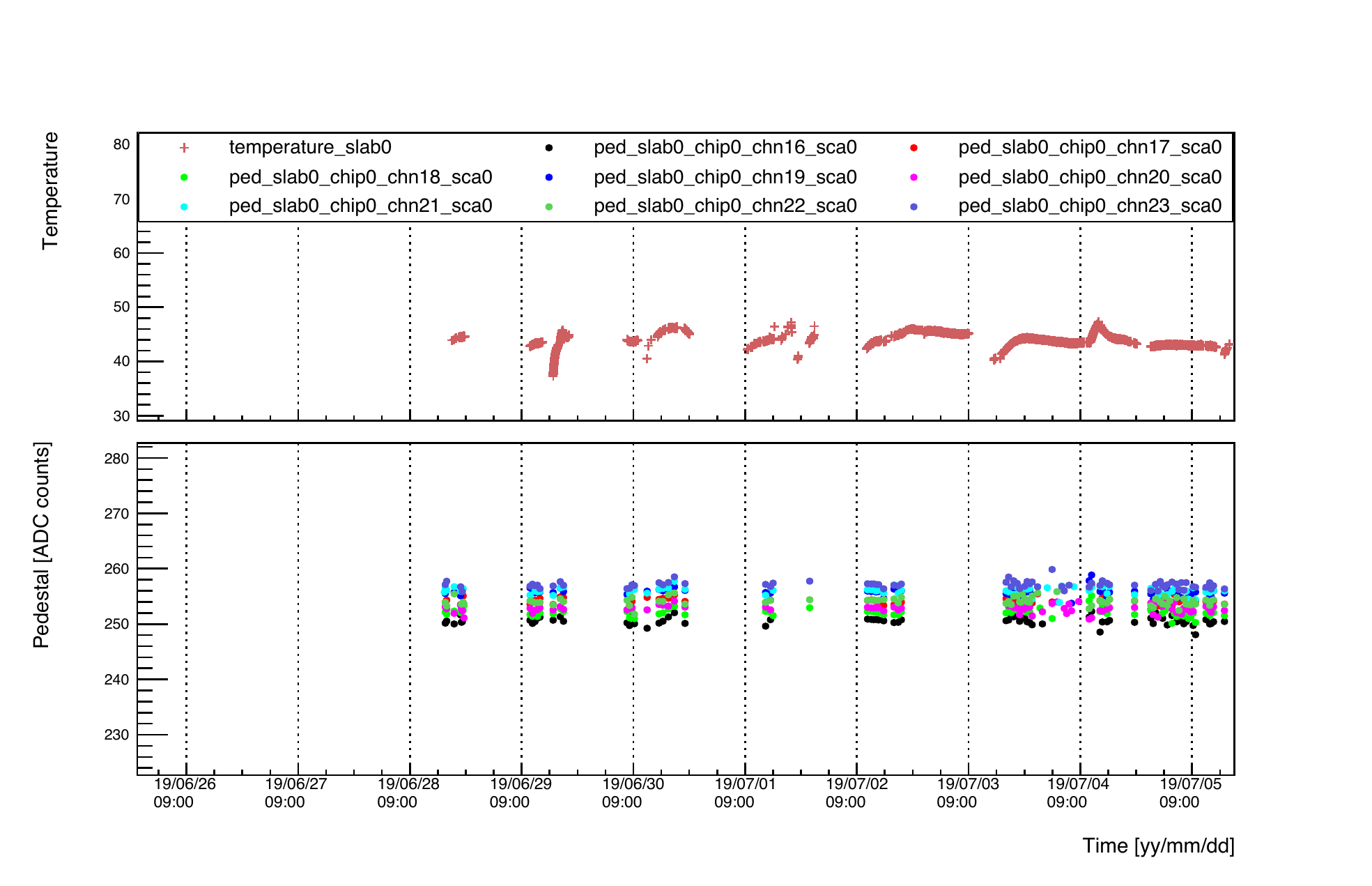}
 \caption{Variation of pedestal means and time dependance of temperature}
 \label{pedestal_temp}
\end{figure}

Figure \ref{slab_temp} and \ref{pedestal_temp} show the analysis results about slabs  during the this test beam.

In Figure \ref{slab_temp}, time dependance of temperature for each slabs is shown.
We can confirm that the temperature of slab becomes higher in the inner part than the outer part of the box.
The dropping part of the temperature show the beginning of the measurements on the day.

In Figure \ref{pedestal_temp}, variations of pedestal means and time dependance of temperature are compared.
Top figure and bottom figure show the time dependance of temperature for slab P1 and the variation of the pedestal means for 8 channels in slab P1, respectively.
We can see the pedestal means are stable within few channels and no strong correlation is seen.
\\
\\

We found five issues in this test.
Three issues arose in the connections.
One issue is loose and unstable connections of HDMI.
The measurements were interrupted due to the issue several times.
There is also an issue on the HV connectors, which broke easily and we had to repair by re-soldering.
Finally, the cable connections are too complicated.
We should connect three cables per one slab in tiny space inside the box.

Remaining two issues are about the readout circuit.
One is ``re-triggering".
The re-triggering is dummy hits after the real hit.
The other is ``double pedestal".
The double pedestal is a phenomenon that two peaks are observed in the ADC distribution of the pedestal.
Now both issues are being studied and we have to solve them.

\section{Summary}

In this paper, we showed improvements and operation tests for DESY test beam 2019.
As the operation tests, we performed measurement of ADC with two radiation sources and calculation of TDC calibration factors by charge injection.
We also showed the quick result of this test beam about pedestal analysis.
From this result, we could confirm that the pedestal is stable during test beam.

As the next step, we have to measure the TDC for all channels and calculate the factors.
We also need to fix the connection problems in the next version of technological prototype.


\end{document}